\documentclass[12pt]{article}
\usepackage{amssymb}
\usepackage{epsfig}
\usepackage{graphicx}
\usepackage{amsmath}
\usepackage{verbatim}

\csname @addtoreset\endcsname{equation}{section}
\begin{document}
\begin{titlepage}
\title{\bf \Large  Electroweak Precision Tests On the MSSM and NMSSM Constrained at the LHC  \vspace{12pt}}

\author{\normalsize Sibo~Zheng  and Yao~Yu \vspace{12pt}\\
{\it\small  Department of Physics, Chongqing University, Chongqing 401331, P.R. China}
\\}

\date{}
\maketitle \voffset -.3in \vskip 1.cm \centerline{\bf Abstract}
\vskip .3cm Following previous works regarding electroweak precision tests in supersymmetry,
 we continue to explore their implications to the MSSM and NMSSM both of which are tightly constrained by
 the Higgs mass and its decay experiments at the LHC.
We adopt universal squark,  slepton mass of three generations,
and A-term of all SM fermions.
We determine the flows of both MSSM and NMSSM in the $S$-$T$ plane with increasing $\tan\beta$ and/or $\gamma\gamma$ rate.
Either increasing the value of $R_{\gamma}$ or $\tan\beta$,
we find that the benchmark points tend to be pushed outside the contour at 99\%\ CL in the MSSM,
however, conversely they tend to shrink towards to the SM reference point in the NMSSM.
In particular, $\tan\beta\gtrsim 50$ is excluded at 99\% CL by the EWPTs for the MSSM due to this behavior.
This behavior is also useful to distinguish the NMSSM from the MSSM.
A byproduct is that the contributions coming from the neutralinos and charginos
should be more suppressed in the MSSM in comparison with the NMSSM.

\vskip 3.cm \noindent March 2013
 \thispagestyle{empty}

\end{titlepage}
\section{Introduction}
Both the ATLAS and CMS collaborations have now established
that the existence of a standard model (SM)-like Higgs with mass resonance at 125 GeV in the $\gamma\gamma$ decay channel \cite{discovery1,discovery2}.
Experimentally it is expected to obtain more precise measurements in this and other decay channels (see, e.g.,\cite{ww}) involving this scalar with higher statistics at the large hadron collider (LHC).

Theoretically, they are various frameworks of new physics in which the present LHC data can be accommodated.
What is of particular interest is the implications to supersymmetric (SUSY) models.
When we consider the minimal supersymmetric standard model (MSSM) \cite{mssm},
large $\tan\beta\gtrsim 30$, heavy stop mass of order $\sim$ several TeVs and large mixing effect with $A$ of order $\sim$ TeV are all needed.
Regardless of mild or severe fine tuning from the concerns of \em{naturalness}\em~\cite{1201.5305} in the MSSM, 
its parameter space is already limited.
Whereas the situation for the next-to- minimal supersymmetric model (NMSSM) \cite{nmssm} is rather different from that of NMSSM 
In this model, small $\tan\beta<10$, large mixing between the singlet scalar and neutral scalars of Higgs two doublets are favored (see, e.g, \cite{1202.5821}). 

The status where both MSSM and NMSSM are tightly constrained by the LHC data reminds us an old tool, i.e,
the electroweak precision tests (EWPTs) \cite{Peskin1,Peskin2},
which can be very useful in the circumstance similar to the MSSM and NMSSM.
Less attention has been paid to applying EWPTs to SUSY models.
The reason was due to the facts that the EWPT is only powerful to constrain new physics that includes few parameters
and that SUSY models possess hundreds of new parameters.

In this paper, we explore the implications of EWPTs on SUSY models.
\em{More constrained the parameter space, more powerful the EWPT}\em.
EWPT is known to play an important role in model building of new physics.
For example, it favors a SM with Higgs mass below 200 GeV and excludes minimal technicolor \cite{technicolor}.
The flaw of old understanding on applying EWPTs to SUSY seems to vanish,
this success might also be achieved in the MSSM \cite{mssm}(and NMSSM \cite{nmssm} as discussed below) tightly constrained by the LHC experiments.

As the favored value of $\tan\beta$ is different between MSSM and NMSSM,
it is expected to see different flows of benchmark points between these two models in the two-parameter plane of $S$ and $T$ of EWPTs.
To see how it flows,
one should obtain the limited parameter spaces of them, respectively.
This can be worked out in terms of the LHC data \cite{discovery1,discovery2,ww}.
Then it is straightforward to extract a set of benchmark points corresponding to
a particular enhancement factor of $\gamma\gamma$ rate, $R_{\gamma}$ relative to SM value,
from the constrained parameter space.
After that we encode the parameters of each benchmark point into the formulae of oblique parameters of MSSM and NMSSM that parameterize the contributions to EWPT observables.
Finally, each benchmark point corresponds to a point
and each curve of $R_{\gamma}$ corresponds to a chain composed of these points in the plane of $S-T$.

Comparing these points with contours set by experiment limits in the plane of $S-T$ (for a review on the experimental status of EWPT, see e.g., \cite{ewptexp}) illustrates whether they are already excluded ?
If not, following the flow of these benchmark points till the one that jumps out the contour at 99\% CL ,
one can determine the limits on the parameters set by EWPTs.

The paper is organized as follows.
First, we briefly discuss the region of parameter space which we would like to scan.
The extraction of parameter space for the MSSM and NMSSM is presented in subsection 2.1 and 2.2 respectively.
The primary results are summarized in Table 2 and 3,
where benchmark points are divided by two quantities,
one the value of $\tan\beta$ and the other the contour of $R_{\gamma}$.
This choice mainly follows from the fact the MSSM and NMSSM are sensitive to different value of $\tan\beta$,
and correlation to the LHC data manifests by the value of $R_{\gamma}$.

Second, we address the contributions to EWPT observables, or equivalently to oblique parameters in the MSSM and NMSSM.
For earlier works, see \cite{Zheng1,Zheng2, mssmewpt1,mssmewpt2, mssmewpt3,mssmewpt4} and discussions presented in section 3.
Combine our previous works \cite{Zheng1,Zheng2}, we present the complete formulae for oblique paramors in appendix A ( viable for large mixing effect ).

Third, in section 3.2 we show the flow of each chain of $R_{\gamma}$ which is composed of a few benchmark points.
It turns out that the $\tan\beta\gtrsim 50$ is excluded at 99\% CL by the EWPTs for the MSSM that predicts a low $R_{\gamma}=1.3$.
Larger value of $R_{\gamma}$, smaller $\tan\beta$ is allowed.

As for the NMSSM, in section 3.3,
we show the flow of each chain of $R_{\gamma}$ also.
It turns out that the fit to the data of EWPT is better given larger $R_{\gamma}$ and $\tan\beta$.
These distinct features are rather useful to distinguish the NMSSM from the MSSM.
A byproduct is that the contributions coming from the neutralinos and charginos
should be more suppressed in the MSSM in comparison with the NMSSM.
We finally conclude in section 4.

\section{The Scan and Parameter Space}
It is convenient to adopt universal squark soft mass $M_{\tilde{Q}}$, and slepton soft mass $M_{\tilde{L}}$ of three generations.
Note that the mass splitting between the two real scalars of either a first two-generation squark or slepton complex scalar is rather small due to the suppression by their small fermion masses.
Therefore,  relaxing the universal assumption to allow the first two generation an common soft mass which however differs from the third generation doesn't give rise to substantial ampliation of the contributions to observables of EWPT.
We also assume universal $A$ term, $A_{\tilde{Q}}=A_{\tilde{L}}=A$ for all SM fermions.
We scan the parameter space in the following regions,
\begin{eqnarray}{\label{E1}}
0&<& m_{\tilde{Q}} < 3~TeV, \nonumber\\
0&<& m_{\tilde{L}} < 3~TeV, \nonumber\\
0&<& A < 3~TeV, \\
0&<& \mu < 3~TeV, \nonumber\\
0&<& \tan\beta < 60. \nonumber
\end{eqnarray}
and additionally,
\begin{eqnarray}{\label{E2}}
0< \lambda < 0.8,~~~~~~~
0< \kappa < 1.
\end{eqnarray}
for NMSSM.
In \eqref{E1}, we allow rather large upper limits of mass scale,
although it may invoke mild or severe fine tuning.
The main reason is partially due to the facts the SM-like Higgs mass of 125 GeV
and enhancement of $R_{\gamma}$ with respective to SM value
have obviously suggested the existence of fine tuning in the MSSM.
In \eqref{E2}, we set the upper limit on $\lambda\sim$0.8,
it follows from considerations of grand unification of SM gauge couplings and Landau pole
(from viewpoint of weak ultra-violet completion).

Apart from the LHC data about the SM-like Higgs experiments,
 we outline in Table 1 the constraints which we would impose on the masses of various supersymmetric
particles from the existing LEP, Tevatron and LHC experiments.
\begin{table}[h]
\centering
\begin{tabular}{|c|c|c|c|c|}
 \hline
  $ m_{\tilde{t}_{1}}\gtrsim 400$ GeV, \cite{stop1,stop2}  \\
  \hline
  $ m_{\tilde{\tau}_{1}}\gtrsim 100$ GeV, \cite{limit1,limit2} \\
  \hline
  $ m_{H^{\pm}}\gtrsim 300$ GeV,  \\
  \hline
  $m_{\tilde{N}_{1}}\gtrsim 100$ GeV, \cite{limit1,limit2}\\
  \hline
  $m_{\tilde{C}^{+}_{1}}\gtrsim 100$ GeV, \cite{limit1,limit2} \\
  \hline
\end{tabular}
\caption{Mass bounds on supersymmetric mass scales as hinted by the LEP, Tevatron and LHC experiments.}
\end{table}

\subsection{Constraint on the MSSM from the LHC}
Now we consider the parameter space constrained by the three requirements, i.e,
$(a)$ the lighter of the two neutral scalar masses is set to be 125 GeV;
$(b)$ the rate ratio of Higgs decay to $\gamma\gamma$ relative to its SM expectation defined as
$R_{\gamma}=\Gamma(h\rightarrow \gamma\gamma)/\Gamma_{SM}(h\rightarrow \gamma\gamma)$ is restricted as $R_{\gamma}=1.8\pm 0.5$ \cite{discovery1,discovery2}.
Here we follow the total ATLAS and CMS combination;
$(c)$ the deviation of rate ratio of h to WW and ZZ to four-lepton channels,
$R_{V}=\Gamma(h\rightarrow VV^{*})/\Gamma_{SM}(h\rightarrow VV^{*})$(V=W, Z),
from its SM expectation is restricted as $R_{W}=1.1\pm 0.3$ \cite{ww}.
Here, we consider the combination of CMS and ATLAS 2012 data.

We present in Table 2 benchmark points $MSSM1$ to $MSSM4$,
which corresponds to $\tan\beta=30$, 40, 50 and 60 in the MSSM, respectively.
In each $MSSMi$ $(i=1,\cdots,4)$,
the left, middle and right column refers to the contour of $R_{\gamma}$ equal to 1.3, 1.8 and 2.0 respectively.
\begin{table}[h!]
\centering
\begin{tabular}{|c|c|c|c|c|c|c|c|c|c|c|c|c|}
 \hline
  & \multicolumn{3}{c|}{$MSSM1(\circ)$} &\multicolumn{3}{c|}{ $MSSM2 (\bullet)$ } &\multicolumn{3}{c|}{ $MSSM3(\star)$ } &\multicolumn{3}{c|}{ $MSSM4(\diamond)$} \\
  \hline
  $\mu$ &2000 &2000 &2000 & 2000& 2000& 2000&2000 & 2000& 2000& 2000& 2000&2000  \\
  \hline
  $M_{\tilde{Q}}$ &928.0 &928.0 &928.0 &928.0 & 928.0&928.0 &928.0 &928.0 &928.0 & 928.0&928.0 &928.0 \\
  \hline
  $M_{\tilde{L}}$ &354.5 &338.3 & 336.2&409.3 & 390.6& 388.3&457.6 &436.7 &434.1 & 501.3&478.4 &475.5 \\
  \hline
  $A$ &2608 & 2608&2608 &2590 & 2590& 2590&2580 &2580 &2580 &2574 &2574 & 2574\\
  \hline
\end{tabular}
\caption{Benchmark Points $MSSM1$ to $MSSM4$ corresponds to $\tan\beta=30$, 40, 50 and 60 in the MSSM, respectively.
The average mass of top scalars is taken as $880$ GeV.
For each $\tan\beta$, the left, middle and right column refers to the contour of $R_{\gamma}$ equal to 1.3, 1.8 and 2.0 respectively.
$<S>$ is in unite of GeV. }
\end{table}

The value of $M_{\tilde{Q}}$ is almost unaltered in Table 2.
To interpret this, note that the Higgs mass $m_{h}$ is sensitive to $\tan\beta$ and  $M_{\tilde{t}}=M_{\tilde{Q}}$.
While $\tan\beta >30$, it rarely affects $m_{h}$,
thus $M_{\tilde{Q}}$ is nearly invariant to guarantee  125 GeV Higgs mass.
The value of $\mu$ is taken to close to its upper limit.
It is a result of the requirement of large deviation to SM coupling,
which is proportional to $\mu\tan\beta$.

\subsection{Constraint on the NMSSM from the LHC}
The situation for the NMSSM is rather different from that of the MSSM.
The tree-level contribution to $m_{h}$ includes an additional term involving Yukawa coupling $\lambda$.
With $\lambda\sim 0.5-0.8$, this term substantially affects the $m_{h}$ dependence on $\tan\beta$,
whose maximal value favors a small $\tan\beta\sim 2-4$.

\begin{table}
\centering
\begin{tabular}{|c|c|c|c|c|c|c|}
 \hline
  & \multicolumn{2}{c|}{$NMSSM1(\circ)$} &\multicolumn{2}{c|}{ $NMSSM2 (\bullet)$ } &\multicolumn{2}{c|}{ $NMSSM3(\star)$ }
 \\
  \hline
$<s>$ & 1499.5 &  1871 & 2430 &  1739& 2160 &  2760  \\
  \hline
$\tan^{2}\beta$ & 5 &  5 & 6 &  6& 7 &  7  \\
\hline
\end{tabular}
\caption{Benchmark Points $NMSSM1$ to $NMSSM3$ which corresponds to $\lambda=0.40$, $\kappa=0.37$, $A_{t}=2647$GeV, $A_{\lambda}=185$ GeV, $A_{\kappa}=-123$ GeV and $m_{\tilde{Q}}$= $m_{\tilde{L}}$ =800 GeV. Here $<s>$ refer to the vacuum expectation value of singlet $S$.
$<S>$ in unite of GeV.}
\end{table}

We present in Table 3 benchmark points $NMSSM1$ to $NMSSM3$,
which corresponds to $\tan^{2}\beta=5$, 6 and 7 in the NMSSM, respectively.
In each $NMSSMi$ $(i=1,\cdots,3)$, similar to Table one,
the left and right column refers to the contour of $R_{\gamma}$ equal to 1.3 and 1.5, respectively.
So far the contributions arising from the fermionic SUSY freedoms such as the neutralinos and charginos have not been included.
Take these effects into account, one expects additional enhancement on either $S$ or $T$.
We leave it for study elsewhere \cite{Zheng4}.

\section{EWPTs }
\subsection{Oblique Parameters at One Loop}
The calculation of oblique parameters given a new physics beyond SM involves loop-induced self energies of SM electroweak gauge bosons.
Following the parametrization of quantum corrections to EWPT observables in \cite{Peskin1, Peskin2},
one can write a set of compact formulae of oblique parameters in terms of the self energies.

For SUSY models that we concern,
the one-loop self-energies in the context of MSSM can be found in earlier works in \cite{mssmewpt1,mssmewpt2}
(see also \cite{mssmewpt3,mssmewpt4}) and reference therein,
and recent considerations in our previous works \cite{Zheng1,Zheng2}.
The compact formulae of the three oblique parameters in the MSSM are also presented in \cite{Zheng1,Zheng2},
while those of NMSSM can be found in appendix A.
An attempt to obtain the results of NMSSM via simulation is shown in \cite{nmssmewpt}.

In appendix A.1, we explicitly present the contribution of squarks and sleptons to the oblique parameters in the case of
mixing between left- and right-hand state.
When these mixing effects are substantial, as hinted by the LHC data in the MSSM,
it must be incorporated.
As the Lagrangian of the squark and slepton states in the MSSM and NMSSM is the same,
the results in appendix A.1 are thus viable for both of them.
Nevertheless, the structure in the Lagrangian for the Higgs sector in these two models is different.
In the NMSSM there is an additional neutral scalar $S$ and neutral fermion recorded in the Higss scalar and neutralino states, respectively.
It turns out that contributions to the oblique parameters from the Higgs sector of the MSSM rather differ from that of NMSSM.
Appendix A.2 involves this issue,
where we present the results that are viable for the NMSSM based on our previous results in \cite{Zheng1}.

The examination on the results in Appendix A is straightforward.
Consider vanishing mixing effects between left- and right-hand state,
the result in appendix A.1 should reduce to those of \cite{Zheng1}.
Similarly it  happens for those in appendix A.2,
 when one considers vanishing singlet states.

\subsection{Implications to the Constrained MSSM From EWPTs}
We can estimate the contribution to $S$ and $T$ for each benchmark point outlined in Table 2.
As mentioned in the introduction,
each benchmark point represents a point in the plane of $S-T$,
and each curve of $R_{\gamma}$ is expressed as a chain of these points.
\begin{figure}[h!]
\centering
\begin{minipage}[b]{0.6\textwidth}
\centering
\includegraphics[width=3.5in]{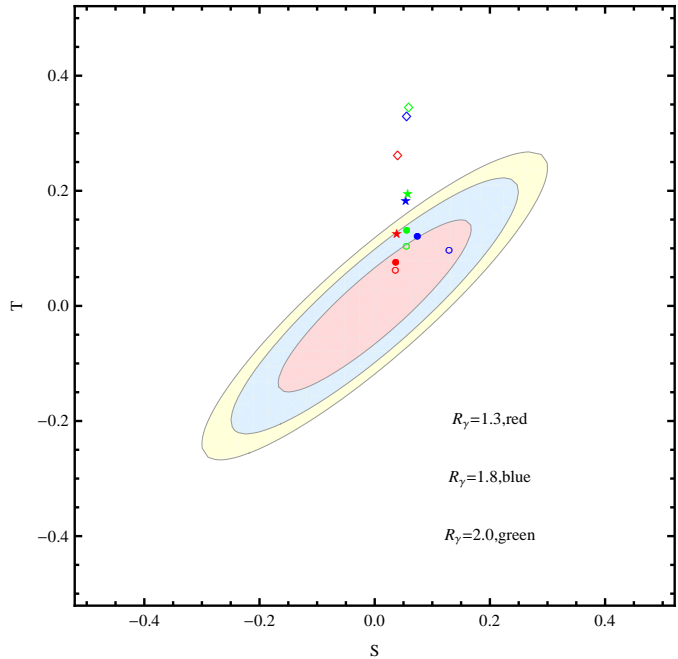}
\end{minipage}%
\caption{Contributions in the benchmark points of Table 2 as indicated in the plane of $S-T$.
We show the three chains that correspond to different value of $R_{\gamma}$.
Given a chain, we find that benchmark point with larger $\tan\beta$
rapidly flows from the SM reference point $(0,0)$ to crossing the contour at 99\% CL. }
\end{figure}

We present the results in Fig.1.
Benchmark points' contribution in the MSSM as indicated in the plane of $S-T$.
We show the three chains with different color correspond to different value of $R_{\gamma}$.
We find that:\\

$(1)$. Consider a chain of same color, benchmark point rapidly flows from the SM reference point $(0,0)$\footnote{This SM reference point corresponds to $m_{h}=125$ GeV and $m_{t}=173$ GeV. }
to crossing the contour at 99\% CL with increasing $\tan\beta$.
This means larger $\tan\beta$ is less favored by the EWPTs.
Typically, $\tan\beta\gtrsim $ 50, 40 and 30 is excluded for $R_{\gamma}$ equal to 1.3, 1.7 and 2.0 respectively.
The large value $R_{\gamma}$, if further favored with higher statistics at the LHC,
will nearly exclude the MSSM at all.

$(2)$. For smaller average stop mass $m_{\tilde{t}}$,
one needs larger mixing effect between left- and right-hand stop scalars so as to explain the 125 GeV mass.
For example, the choice $m_{\tilde{t}}=770$ GeV instead of 880 GeV in Table 2 requires $A^{2}_{t}/m^{2}_{\tilde{t}}\simeq 6.5$,
which is closer to the maximal mixing.
Larger mixing gives rise to more obvious mass splittings between left- and right-hand states,
consequently larger contributions to $S$ and $T$.
We illustrate in Fig. 2 what changes for this replacement.
In comparison with Fig.1, there are indeed more points pushed out the contour at 99\% CL.
With larger $R_{\gamma}$, more substantial this modification is.
Therefore, large $R_{\gamma}$ favors small mixing effect and large $m_{\tilde{t}}$.

(3). Take the contributions to $S$ and $T$ from the neutralinos and charginos into account,
it should not change much, or there exists a subtle cancelation among these additional effects.
Otherwise, the MSSM is nearly excluded by present LHC data.
This issue will be discussed elsewhere \cite{Zheng4}.
\begin{figure}[h!]
\centering
\begin{minipage}[b]{0.6\textwidth}
\centering
\includegraphics[width=3.5in]{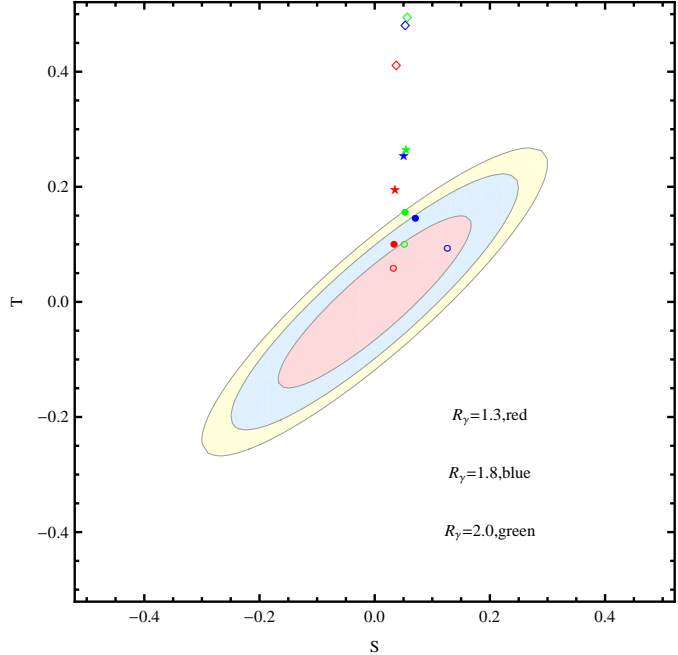}
\end{minipage}%
\caption{Benchmark points' contributions in the MSSM as indicated in the plane of $S-T$,
with the average mass of top scalars taken as $770$ GeV.
In this case $A\sim 2.0$ TeV, which is smaller than that in Table 2.}
\end{figure}

\subsection{Implications to the Constraints NMSSM From EWPTs}
Now we discuss the contribution to $S$ and $T$ for each benchmark point as shown in Table 3.
Similar to the previous subsection,
each benchmark point corresponds to a point in the plane of $S-T$,
and each contour line of $R_{\gamma}$ is expressed as a chain of these points.

The result is shown in Fig.3,
where contours with different colors indicate different $R_{\gamma}$.
In contrast to observations $(1)$ and $(2)$ in the MSSM,
we observe that,\\

$(1)$, Compare the blue and the red contour in Fig.3,
one finds that for larger $R_{\gamma}$,
 the fit to the data of EWPT is better.
Because those points tend to retain at the reference point.
It follows from this observation that large $\gamma\gamma$ rate  favors NMSSM other than MSSM.

$(2)$,  Moreover, for either the red or blue contour,
one finds that  increasing $\tan\beta$ results in the suppression of the contribution.
This distinct feature can be useful to distinguish the NMSSM from the MSSM.

$(3)$, The fit to the data of EWPTs is rather good.
This implies that the contributions coming from the neutralinos and charginos should be more suppressed in the MSSM in comparison with the NMSSM,
unless a subtle cancelation exists as mentioned.
In this sense, the MSSM seems more fine tuned.
\begin{figure}[h!]
\centering
\begin{minipage}[b]{0.6\textwidth}
\centering
\includegraphics[width=3.5in]{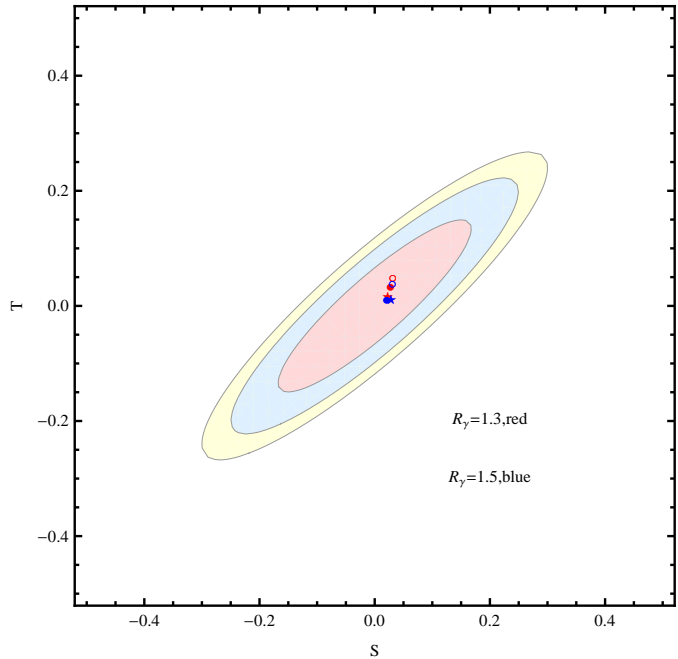}
\end{minipage}%
\caption{ Contributions arising from benchmark points $NMSSM1$ to $NMSSM3$ is shown in the plane of $S-T$. Each chain corresponds to a fixed value of $R_{\gamma}$. Given a chain, we find that benchmark point corresponding to either larger $\tan\beta$ or $R_{\gamma}$ tends to flow to the SM reference point $(0,0)$. }
\end{figure}

\section{Conclusions}
We discuss the impliations of the LHC and EWPT experiments together to MSSM and NMSSM.
The flow of each chain of $R_{\gamma}$ that corresponds to a few benchmark points is explicitly shown.
In the plane of $S-T$, we observe that either increasing the value of $R_{\gamma}$ or $\tan\beta$,
the benchmark points tends to be pushed outside the contour at 99\%\ CL in the MSSM,
however, conversely they tends to shrink in the NMSSM.
The fit to the EWPT data is better in the NMSSM other than the MSSM due to this behavior.
In particular, $\tan\beta\gtrsim 50$ is excluded at 99\% CL by the EWPTs for the MSSM.
Also this behavior is rather useful to distinguish the NMSSM from the MSSM.
A byproduct is that the contributions coming from the neutralinos and charginos
should be more suppressed in the MSSM in comparison with the NMSSM.

It would be interesting to address the NMSSM with large $\lambda$,
which is usually induced at low energy by strongly coupled dynamics at high energy region.
The study will be useful to answer the question whether either weak or strong UV theory is more favored by the present experiments.

Our results are preliminary.
As we only focus our attention to the scalar SUSY freedoms,
the effects arising from neutralinos and charginos should be taken into account for completion.
Also the benchmark points are not large enough.
It will be of interest to estimate modifications to our observations when one considers a set of larger samples. \\

~~~~~~~~~~~~~~~~~~~~~~~~~~~~~~~~~~~~~~~~
$\bf{Acknowledgement}$\\
We are grateful to M-x, Luo and Tianjun Li for reading the manuscript and valuable comments.
This work is supported by the Natural Science
Foundation of China under Grant No. 11247031.\\

\appendix
\section{Oblique Parameters}
Following the same method as in our previous works \cite{Zheng1,Zheng2},
we derive the $S$, $T$ ans $U$ parameter viable for the NMSSM.
\subsection{ Sfermion Sector with Mass Mixing}
The squarks of each generation give rise to the $S$, $T$ as follows,
\begin{eqnarray}
S&=&-\frac{4N_{c}}{\pi}\left\{s^{2}c^{2}\left[\frac{4}{9}B^{'}_{22}(0,m^{2}_{\widetilde{u}_{1}},m^{2}_
{\widetilde{u}_{1}})+\frac{4}{9}B^{'}_{22}(0,m^{2}_{\widetilde{u}_{2}},m^{2}_
{\widetilde{u}_{2}})+\frac{1}{9}B^{'}_{22}(0,m^{2}_{\widetilde{d}_{1}},m^{2}_
{\widetilde{d}_{1}})+\frac{4}{9}B^{'}_{22}(0,m^{2}_{\widetilde{d}_{2}},m^{2}_
{\widetilde{d}_{2}})\right]\right.\nonumber\\
&    &\left.-\left[\frac{1}{4}(g_{u_{L}}+g_{u_{R}})^{2}B^{'}_{22}(0,m^{2}_{\widetilde{u}_{1}},m^{2}_
{\widetilde{u}_{1}})+\frac{1}{4}(g_{u_{L}}+g_{u_{R}})^{2}B^{'}_{22}(0,m^{2}_{\widetilde{u}_{2}},m^{2}_
{\widetilde{u}_{2}})\right.\right.\nonumber\\
&    &\left.\left.+\frac{1}{2}(g_{u_{L}}-g_{u_{R}})^{2}B^{'}_{22}(0,m^{2}_{\widetilde{u}_{1}},m^{2}_
{\widetilde{u}_{2}})+\frac{1}{4}(g_{d_{L}}+g_{d_{R}})^{2}B^{'}_{22}(0,m^{2}_{\widetilde{d}_{1}},m^{2}_
{\widetilde{d}_{1}})\right.\right.\nonumber\\
&    &\left.\left.+\frac{1}{4}(g_{d_{L}}+g_{d_{R}})^{2}B^{'}_{22}(0,m^{2}_{\widetilde{d}_{2}},m^{2}_
{\widetilde{d}_{2}})+\frac{1}{2}(g_{d_{L}}-g_{d_{R}})^{2}B^{'}_{22}(0,m^{2}_{\widetilde{d}_{1}},m^{2}_
{\widetilde{d}_{2}})\right]\right.\nonumber\\
&   &+\left.(c^{2}-s^{2})\left[\frac{1}{3}(g_{u_{L}}+g_{u_{R}})B^{'}_{22}(0,m^{2}_{\widetilde{u}_{1}},m^{2}_{\widetilde{u}_{1}})+
\frac{1}{3}(g_{u_{L}}+g_{u_{R}})B^{'}_{22}(0,m^{2}_{\widetilde{u}_{2}},m^{2}_{\widetilde{u}_{2}})\right.\right.\nonumber\\
&    &\left.\left.+\frac{1}{3}(g_{d_{L}}+g_{d_{R}})B^{'}_{22}(0,m^{2}_{\widetilde{d}_{1}},m^{2}_{\widetilde{d}_{1}})+
\frac{1}{3}(g_{d_{L}}+g_{d_{R}})B^{'}_{22}(0,m^{2}_{\widetilde{d}_{2}},m^{2}_{\widetilde{d}_{2}})\right]
\right\}\nonumber\\
\end{eqnarray}
and
\begin{eqnarray}
T&=&\frac{N_{c}}{4\pi s^{2}m^{2}_{W}}\left\{\frac{1}{2}B_{22}(0,m^{2}_{\widetilde{u}_{1}},m^{2}_{\widetilde{d}_{1}})+
\frac{1}{2}B_{22}(0,m^{2}_{\widetilde{u}_{1}},m^{2}_{\widetilde{d}_{2}})
+\frac{1}{2}B_{22}(0,m^{2}_{\widetilde{d}_{1}},m^{2}_{\widetilde{u}_{2}})
+\frac{1}{2}B_{22}(0,m^{2}_{\widetilde{u}_{1}},m^{2}_{\widetilde{u}_{2}})\right.\nonumber\\
&    &\left.-\frac{1}{4}A_{0}(m_{\widetilde{u}_{1}})-\frac{1}{4}A_{0}(m_{\widetilde{u}_{2}})-
\frac{1}{4}A_{0}(m_{\widetilde{d}_{1}})-\frac{1}{4}A_{0}(m_{\widetilde{d}_{2}})\right.\nonumber\\
&    &\left.-\left((g_{u_{L}}+g_{u_{R}})^{2}B_{22}(0,m^{2}_{\widetilde{u}_{1}},m^{2}_
{\widetilde{u}_{1}})+(g_{u_{L}}+g_{u_{R}})^{2}B_{22}(0,m^{2}_{\widetilde{u}_{2}},m^{2}_
{\widetilde{u}_{2}})+2(g_{u_{L}}-g_{u_{R}})^{2}B_{22}(0,m^{2}_{\widetilde{u}_{1}},m^{2}_
{\widetilde{u}_{2}})\right.\right.\nonumber\\
&    &\left.\left.(g_{d_{L}}+g_{d_{R}})^{2}B_{22}(0,m^{2}_{\widetilde{d}_{1}},m^{2}_
{\widetilde{d}_{1}})+(g_{d_{L}}+g_{d_{R}})^{2}B_{22}(0,m^{2}_{\widetilde{d}_{1}},m^{2}_
{\widetilde{d}_{2}})+2(g_{d_{L}}-g_{d_{R}})^{2}B_{22}(0,m^{2}_{\widetilde{d}_{1}},m^{2}_
{\widetilde{d}_{2}})\right.\right.\nonumber\\
&    &\left.\left.-({g_{u_{L}}}^{2}+{g_{u_{R}}}^{2})A_{0}(m_{\widetilde{u}_{1}})-({g_{u_{L}}}^{2}+{g_{u_{R}}}^{2})A_{0}(m_{\widetilde{u}_{2}})
-({g_{d_{L}}}^{2}+{g_{d_{R}}}^{2})A_{0}(m_{\widetilde{d}_{1}})\right.\right.\nonumber\\
&    &\left.\left.-({g_{d_{L}}}^{2}+{g_{d_{R}}}^{2})A_{0}(m_{\widetilde{d}_{2}})
\right)
-2s^{2}\left[\frac{4}{3}(g_{u_{L}}+g_{u_{R}})B_{22}(0,m^{2}_{\widetilde{u}_{1}},m^{2}_
{\widetilde{u}_{1}})+\frac{4}{3}(g_{u_{L}}+g_{u_{R}})B_{22}(0,m^{2}_{\widetilde{u}_{2}},m^{2}_{\widetilde{u}_{2}})\right.\right.\nonumber\\
&    &\left.\left.-\frac{2}{3}(g_{u_{L}}+g_{u_{R}})A_{0}(m_{\widetilde{u}_{1}})
-\frac{2}{3}(g_{u_{L}}+g_{u_{R}})A_{0}(m_{\widetilde{u}_{2}})-\frac{2}{3}(g_{d_{L}}+g_{d_{R}})B_{22}(0,m^{2}_{\widetilde{d}_{1}},m^{2}_
{\widetilde{d}_{1}})\right.\right.\nonumber\\
&    &\left.\left.-\frac{2}{3}(g_{d_{L}}+g_{d_{R}})B_{22}(0,m^{2}_{\widetilde{d}_{2}},m^{2}_{\widetilde{d}_{2}})
-\frac{1}{3}(g_{d_{L}}+g_{d_{R}})A_{0}(m_{\widetilde{d}_{1}})
-\frac{1}{3}(g_{d_{L}}+g_{d_{R}})A_{0}(m_{\widetilde{d}_{2}})\right]
\right\}\nonumber\\
\end{eqnarray}
The sleptons in each generation give rises to
\begin{eqnarray}
S&=&-\frac{4}{\pi}\left\{s^{2}c^{2}(B^{'}_{22}(0,m^{2}_{\widetilde{e}_{1}},m^{2}_
{\widetilde{e}_{1}})+B^{'}_{22}(0,m^{2}_{\widetilde{e}_{2}},m^{2}_
{\widetilde{e}_{2}}))\right.\nonumber\\
&    &\left.-\left[g_{\nu_{L}}^{2}B^{'}_{22}(0,m^{2}_{\widetilde{\nu}_{L}},m^{2}_
{\widetilde{\nu}_{L}})+\frac{1}{4}(g_{e_{L}}+g_{e_{R}})^{2}B^{'}_{22}(0,m^{2}_{\widetilde{e}_{1}},m^{2}_
{\widetilde{e}_{1}})\right.\right.\nonumber\\
&    &\left.\left.+\frac{1}{4}(g_{e_{L}}+g_{e_{R}})^{2}B^{'}_{22}(0,m^{2}_{\widetilde{e}_{2}},m^{2}_
{\widetilde{e}_{2}})+\frac{1}{2}(g_{e_{L}}-g_{e_{R}})^{2}B^{'}_{22}(0,m^{2}_{\widetilde{e}_{1}},m^{2}_
{\widetilde{e}_{2}})\right]\right.\nonumber\\
&    &+\left.(c^{2}-s^{2})\left[
-\frac{1}{2}(g_{e_{L}}+g_{e_{R}})B^{'}_{22}(0,m^{2}_{\widetilde{e}_{1}},m^{2}_{\widetilde{e}_{1}})
-\frac{1}{2}(g_{e_{L}}+g_{e_{R}})B^{'}_{22}(0,m^{2}_{\widetilde{e}_{2}},m^{2}_{\widetilde{e}_{2}})\right]
\right\}\nonumber\\
\end{eqnarray}
and
\begin{eqnarray}
T&=&\frac{1}{4\pi s^{2}m^{2}_{W}}\left[B_{22}(0,m^{2}_{\widetilde{\nu}_{L}},m^{2}_{\widetilde{e}_{1}})+
B_{22}(0,m^{2}_{\widetilde{\nu}_{L}},m^{2}_{\widetilde{e}_{2}})
\right.\nonumber\\
&    &\left.-\frac{1}{2}A_{0}(m_{\widetilde{e}_{1}})-
\frac{1}{4}A_{0}(m_{\widetilde{e}_{1}})-\frac{1}{4}A_{0}(m_{\widetilde{e}_{2}})\right.\nonumber\\
&    &\left.-\left(4g_{\nu_{L}}^{2}B_{22}(0,m^{2}_{\widetilde{\nu}_{L}},m^{2}_
{\widetilde{\nu}_{L}})+(g_{e_{L}}+g_{e_{R}})^{2}B_{22}(0,m^{2}_{\widetilde{e}_{1}},m^{2}_
{\widetilde{e}_{1}})\right.\right.\nonumber\\
&    &\left.\left.+(g_{e_{L}}+g_{e_{R}})^{2}B_{22}(0,m^{2}_{\widetilde{e}_{1}},m^{2}_
{\widetilde{e}_{2}})+2(g_{e_{L}}-g_{e_{R}})^{2}B_{22}(0,m^{2}_{\widetilde{e}_{1}},m^{2}_
{\widetilde{e}_{2}})\right.\right.\nonumber\\
&    &\left.\left.-(2{g_{\nu_{L}}}^{2})A_{0}(m_{\widetilde{\nu}_{L}})
-({g_{e_{L}}}^{2}+{g_{e_{R}}}^{2})A_{0}(m_{\widetilde{e}_{1}})-({g_{e_{L}}}^{2}+{g_{e_{R}}}^{2})A_{0}(m_{\widetilde{e}_{2}})\right)\right.\nonumber\\
&    &\left.-2s^{2}\left(
-2(g_{e_{L}}+g_{e_{R}})B_{22}(0,m^{2}_{\widetilde{e}_{2}},m^{2}_{\widetilde{e}_{2}})
-2(g_{e_{L}}+g_{e_{R}})B_{22}(0,m^{2}_{\widetilde{e}_{2}},m^{2}_{\widetilde{e}_{2}})\right.\right.\nonumber\\
&    &\left.\left.+(g_{e_{L}}+g_{e_{R}})A_{0}(m_{\widetilde{e}_{1}})
+(g_{e_{L}}+g_{e_{R}})A_{0}(m_{\widetilde{e}_{2}})\right)
\right]
\end{eqnarray}
Here $\tilde{u}_{1,2}$, $\tilde{d}_{1,2}$ and $\tilde{e}_{1,2}$ refer to their mass eigenstates,
and quantities $g_{i}=T^{3}_{i}-s^{2}Q_{i}$ with $Q_{i}$ the electric charge and $T^{3}_{i}$ the isospin component of  SM left- and right-hand fermion $\psi_i$.
For details on functionals $B_i$, see, e.g, Refs. \cite{Zheng1,Zheng2}.

\subsection{The Higgs Sector in the NMSSM}
The Higgs sector of the NMSSM contains five physical neutral Higgs bosons,
three Higgs scalars  $S_{a}$ $(a = 1, 2, 3)$ and two pseudoscalars $P_{\alpha}$ $(\alpha = 1, 2)$,
and two degenerate physical charged Higgs particles $C_{\pm}$\footnote{We follow the convention and notations for the Higgs sector of NMSSM in Ref. \cite{9512366}. }.
\begin{eqnarray}
S&=&-\frac{1}{\pi}\left[B^{'}_{22}(0,m^{2}_{C^{+}},m^{2}_{C^{+}})-(U^{s}_{a1}U^{p}_{\alpha1}-U^{s}_{a2}U^{p}_{\alpha2})^{2}B^{'}_{22}(0,m^{2}_{s_{a}},m^{2}_{p
_{\alpha}})+B^{'}_{22}(0,m^{2}_{h_{0}},m^{2}_{G_{0}})\right.\nonumber\\
&    &\left.+(U^{s}_{a1}\cos\beta+U^{s}_{a2}\sin\beta)^{2}m^{2}_{Z}B^{'}_{0}(0,m^{2}_{Z},m^{2}_{s
_{a}})-m^{2}_{Z}B^{'}_{0}(0,m^{2}_{h_{0}},m^{2}_{Z})\right]
\end{eqnarray}
\begin{eqnarray}
T&=&-\frac{1}{4\pi s^{2}m^{2}_{W}}\left[(U^{s}_{a1}\sin\beta-U^{s}_{a2}\cos\beta)^{2}B_{22}(0,m^{2}_{C^{+}},m^{2}_{s_{a}})+
(U^{s}_{a1}\cos\beta+U^{s}_{a2}\sin\beta)^{2}B_{22}(0,m^{2}_{G^{+}},m^{2}_{s_{a}})\right.\nonumber\\
&    &\left.-B_{22}(0,m^{2}_{G^{+}},m^{2}_{h_{0}})+(U^{p}_{\alpha1}\sin\beta+U^{p}_{\alpha2}\cos\beta)^{2}B_{22}(0,m^{2}_{C^{+}},m^{2}_{p_{\alpha}})
-B_{22}(0,m^{2}_{G^{+}},m^{2}_{G_{0}})\right.\nonumber\\
&    &\left.+(U^{p}_{\alpha1}\cos\beta-U^{p}_{\alpha2}\sin\beta)^{2}B_{22}(0,m^{2}_{G^{+}},m^{2}_{p_{\alpha}})-\frac{1}{2}A_{0}(m_{C^{+}})\right.\nonumber\\
&    &\left.-(U^{s}_{a1}U^{p}_{\alpha1}-U^{s}_{a2}U^{p}_{\alpha2})^{2}B_{22}(0,m^{2}_{s_{a}},m^{2}_{p_{\alpha}})
+B_{22}(0,m^{2}_{h_{0}},m^{2}_{G_{0}})\right.\nonumber\\
&    &\left.-(U^{s}_{a1}\cos\beta+U^{s}_{a2}\sin\beta)^{2}
(m^{2}_{W}B_{0}(0,m^{2}_{s_{a}},m^{2}_{W^{+}})-m^{2}_{Z}B_{0}(0,m^{2}_{s_{a}},m^{2}_{Z_{0}}))\right.\nonumber\\
&    &\left.+m^{2}_{W}B_{0}(0,m^{2}_{h_{0}},m^{2}_{W^{+}})-m^{2}_{Z}B_{0}(0,m^{2}_{h_{0}},m^{2}_{Z_{0}})\right]
\end{eqnarray}
\begin{eqnarray}
U&=&\frac{1}{\pi}\left[(U^{s}_{a1}\sin\beta-U^{s}_{a2}\cos\beta)^{2}B^{'}_{22}(0,m^{2}_{C^{+}},m^{2}_{s_{a}})+
(U^{s}_{a1}\cos\beta+U^{s}_{a2}\sin\beta)^{2}B^{'}_{22}(0,m^{2}_{G^{+}},m^{2}_{s_{a}})\right.\nonumber\\
&    &\left.-B^{'}_{22}(0,m^{2}_{G^{+}},m^{2}_{h_{0}})+(U^{s}_{\alpha1}\sin\beta+U^{s}_{\alpha2}\cos\beta)^{2}B^{'}_{22}(0,m^{2}_{C^{+}},m^{2}_{p_{\alpha}})
-B^{'}_{22}(0,m^{2}_{G^{+}},m^{2}_{G_{0}})\right.\nonumber\\
&    &\left.+(U^{s}_{\alpha1}\cos\beta-U^{s}_{\alpha2}\sin\beta)^{2}B^{'}_{22}(0,m^{2}_{G^{+}},m^{2}_{p_{\alpha}})
-B^{'}_{22}(0,m^{2}_{C^{+}},m^{2}_{C^{+}})\right.\nonumber\\
&    &\left.-(U^{s}_{a1}U^{p}_{\alpha1}-U^{s}_{a2}U^{p}_{\alpha2})^{2}B^{'}_{22}(0,m^{2}_{s_{a}},m^{2}_{p_{\alpha}})
+B^{'}_{22}(0,m^{2}_{h_{0}},m^{2}_{G_{0}})\right.\nonumber\\
&    &\left.-(U^{s}_{a1}\cos\beta+U^{s}_{a2}\sin\beta)^{2}
(m^{2}_{W}B^{'}_{0}(0,m^{2}_{s_{a}},m^{2}_{W^{+}})-m^{2}_{Z}B^{'}_{0}(0,m^{2}_{s_{a}},m^{2}_{Z_{0}}))\right.\nonumber\\
&    &\left.+m^{2}_{W}B^{'}_{0}(0,m^{2}_{h_{0}},m^{2}_{W^{+}})-m^{2}_{Z}B^{'}_{0}(0,m^{2}_{h_{0}},m^{2}_{Z_{0}})\right]
\end{eqnarray}
where  $U^{s}$ and $U^{p}$, which are real orthogonal $3\times 3$ matrices,
is used to diagonalize the CP-even mass matrx $\mathcal{M}_{S}$ and CP-odd $\mathcal{M}_{P}$
respectively.

\end{document}